\documentclass[aps,prl,twocolumn,superscriptaddress]{revtex4-2}
\usepackage{amsfonts}
\usepackage{amsmath}
\usepackage{amssymb}
\usepackage{graphicx}
\usepackage{bm}
\usepackage{xcolor}

\renewcommand{\d}{\text{d}}

\begin{document}

\title{Modification of the fluctuation dynamics of ultra-thin wetting films}
\author{C. Clavaud}
\affiliation{CNRS, Sciences et Ing\'enierie de la Mati\`ere Molle, ESPCI Paris, PSL Research University, Sorbonne Universit\'e, 75005 Paris, France}
\author{M. Maza-Cuello}
\affiliation{CNRS, Sciences et Ing\'enierie de la Mati\`ere Molle, ESPCI Paris, PSL Research University, Sorbonne Universit\'e, 75005 Paris, France}
\author{C. Fr\'etigny}
\affiliation{CNRS, Sciences et Ing\'enierie de la Mati\`ere Molle, ESPCI Paris, PSL Research University, Sorbonne Universit\'e, 75005 Paris, France}
\author{L. Talini}
\email[]{laurence.talini@espci.fr}
\affiliation{CNRS, Surface du Verre et Interfaces, Saint-Gobain, 93300 Aubervilliers, France}
\author{T. Bickel}
\affiliation{Univ. Bordeaux, CNRS, Laboratoire Ondes et Mati\`ere d'Aquitaine (UMR 5798), F-33400 Talence, France}

\begin{abstract}
We report on the effect of intermolecular forces on the fluctuations of supported liquid films. Using an optically-induced thermal gradient, we form nanometer-thin films of wetting liquids on glass substrates, where van der Waals forces are balanced by thermocapillary forces. We show that the fluctuation dynamics of the film interface is strongly modified by intermolecular forces at lower frequencies. Data spanning three frequency decades are in excellent agreement with theoretical predictions accounting for van der Waals forces. Our results emphasize the relevance of  intermolecular forces on thermal fluctuations when fluids are confined at the nanoscale. 
\end{abstract}

\maketitle

When shared with air or another fluid, a liquid interface appears macroscopically flat but is actually corrugated by thermal motion at smaller scales. Although of subnanometric amplitude, thermal capillary waves  play a central role in phenomena such as drop formation~\cite{hennequinPRL2006}, film break-up~\cite{chatzigiannakisPRL2020}, jet destabilisation~\cite{moselerScience2000,petitPNAS2012}, as well as wetting~\cite{davidovitchPRL2005,fernandezJCIS2019} or dewetting~\cite{fetzerPRL2007} of liquid films on solid substrates. As first described by Vrij~\cite{vrijFaraday1966} and Sheludko~\cite{sheludkoACIS1967}, the rupture of a non-draining liquid film originates from the amplification of interfacial fluctuations with wavelengths larger than a critical value, which is set by the balance between surface tension and intermolecular forces. The lifetime of metastable supported thin liquid films is therefore ruled by the thermal roughness of their interface. In turn, fluctuations are themselves modified as the thickness of the liquid layer decreases since confinement enhances hydrodynamic dissipation~\cite{jackleJPCM1998,henlePRE2007}. 

Intermolecular forces are also expected to have a significant effect on the fluctuations of ultra-thin films. Indeed, dipolar interactions between the solid substrate, the liquid layer and the gas phase become relevant when the thickness of the film is smaller than a few tens of nanometers~\cite{bickelEPL2014}. The resulting van der Waals forces can be either attractive or repulsive. In the case of attractive interactions, the fluctuations are amplified by intermolecular forces and the film is predicted to be unstable~\cite{meckeJPCM2005,macdowellPRL2013,zhangPRE2019,zhangJFM2021}. This is confirmed by experimental observations on dewetting polymer films of nanometric thickness~\cite{fetzerPRL2007}. Still,  in this nonequilibrium situation, measurements are  limited to the growth of the more unstable mode and the full spectrum of thermal capillary waves remains unexplored. In contrast, the fluctuation amplitude of a liquid layer with repulsive liquid/air and liquid/solid interactions is expected to decrease under the action of intermolecular forces. This qualitatively explains the formation and stability of ultra-thin films of wetting liquids, such as the precursor films involved in the spreading of a macroscopic droplet. Nevertheless, the effect of van der Waals forces on interfacial fluctuation has never been evidenced experimentally. More generally, investigations regarding thermal capillary waves at the interface of nanoscale liquid films remain almost exclusively theoretical. 

Thermal fluctuations of liquid interfaces were first characterized more than forty years ago using light scattering techniques~\cite{bouchiatJPhys1971}. They proved to be an accurate and non-invasive probe of the macroscopic properties of liquids, \textit{e.g.}, surface tension or viscoelasticity, that can be deduced from the capillary wave spectrum for an infinite medium~\cite{langevinbook}. Owing to experimental limitations, measurements have long been limited to very thick liquid layers. 
The effect of confinement has been addressed more recently thanks to the development of novel techniques based on either the deflection of a reflected laser beam~\cite{tayRSI2008,mitsuiPRE2013} or x-ray scattering~\cite{huPRE2006,alvinePRL2012}. Both techniques differ by the range of the wavevectors that can be probed in the experiments. Small wavelength x-ray measurements are better suited for highly viscous fluids such as high molecular weight  polymer melts, whereas laser deflection can be used for simple, low-viscosity liquids. 
Enhancement of experimental accuracy now makes it possible to investigate the modification of the fluctuation dynamics in very thin films~\cite{alvinePRL2012,pottierPRE2014}. Hydrodynamic boundary conditions at the liquid/solid interface were also successfully tested~\cite{pottierPRL2015}. 

However, the contribution of intermolecular forces on the capillary spectrum has never been evidenced yet. The experimental challenges are multiple since stable films of controlled nanometric thickness must be produced in the first place. Then,  interfacial deformations have to be probed at the molecular scales and over long times. To achieve these goals, we make use of the thermocapillary effect to create ultra-thin films of wetting liquids. We then measure the fluctuation dynamics of the interface in a large frequency range using the previously developed Surface Fluctuation Specular Reflection  technique~\cite{tayRSI2008}. We further compare the experimental data with  theoretical predictions based on the lubrication approximation. Our analysis therefore shows  that the thickness of the liquid layer  results from a balance between thermocapillary and intermolecular forces, and that the fluctuation spectra are dominated by van der Waals forces at low frequencies.

%%%%%%%%%%%%%%%%%%%%%%%%%%%%%%%%%%%%%%%
\begin{figure}
\includegraphics[width=0.98\columnwidth]{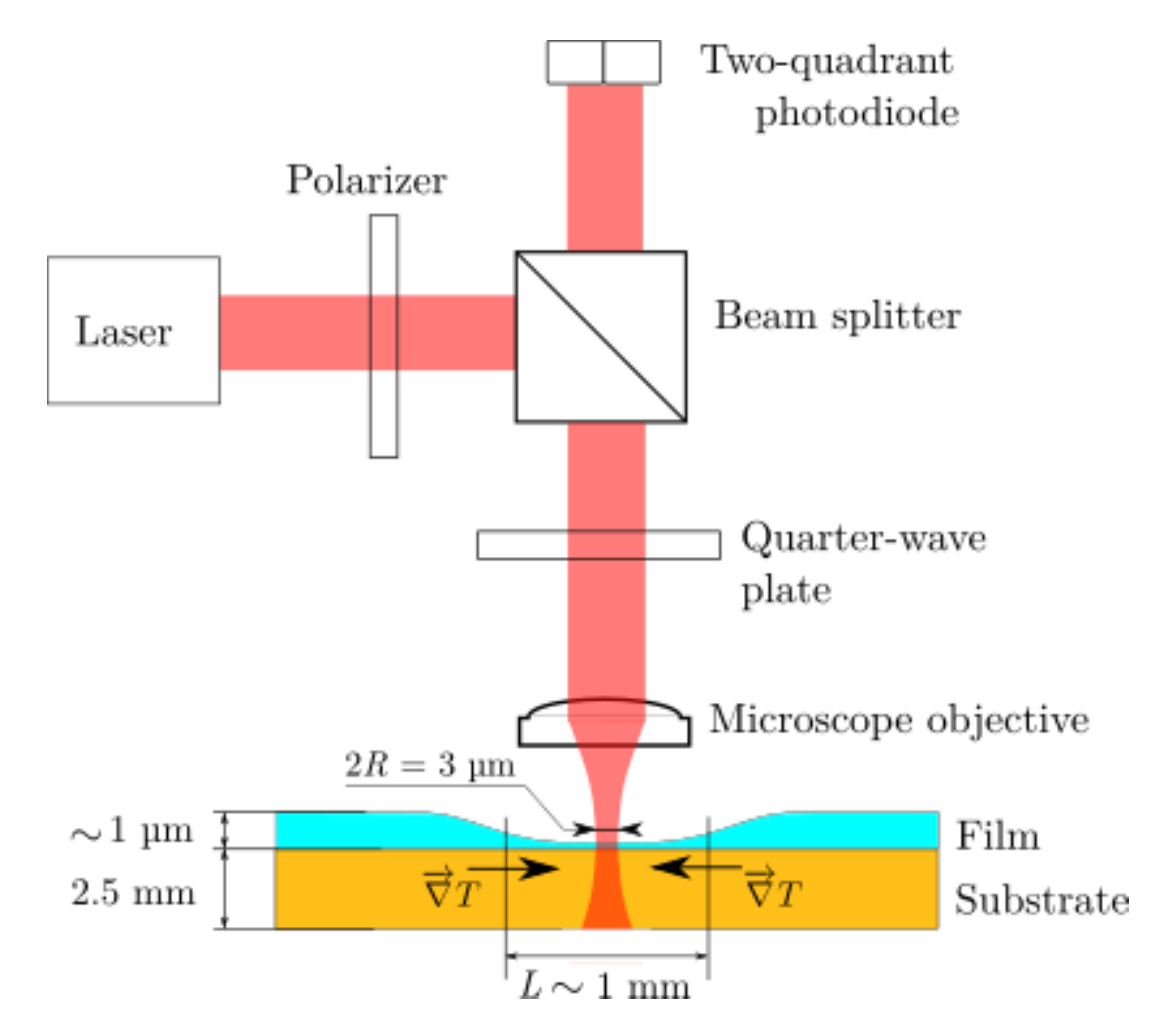}
\caption{Schematized experimental set-up (not to scale). The laser is focused on the liquid film spread on a colored glass substrate. Light absorption by the substrate creates a thermal gradient in the liquid layer over the lengthscale~$L$, which also define the extension of the film thinning.}
\label{fig1}
\end{figure}
%%%%%%%%%%%%%%%%%%%%%%%%%%%%%%%%%%%%%%%

Silicone oils (Rhodorsil) of two different viscosities ($9.3$ and $19$~mPa$\cdot$s at room temperature) are spin-coated on colored glass substrates (Hoya Longpass optical filters, diameter $50$~mm, thickness 2.5~mm) to form films of uniform initial thickness, between 250 and 1150~nm. The orange color of the glass substrate (optical index $n=1.526$) is chosen for its poor but non-zero absorption of the He-Ne laser light (wavelength 632.8~nm). The spin-coated films are placed in the set-up described in Fig.~\ref{fig1}. The laser is strongly focused on the film with a $\times 50$ microscope objective. The other optical elements are used to avoid back laser light and to collect the light reflected by the interface on a two-quadrant photodiode. Special care is taken to suppress parasitic light on the photodiode and to place the liquid film at the laser beam waist within the Rayleigh length ($\approx 20~\mu$m). The reflected beam is centered on the two quadrants and the variations of its position are detected by computing the Power Spectrum Density (PSD) of the difference between the voltages delivered by the quadrants.  Owing to either low or high-frequency parasitic noise, fluctuation spectra are obtained for frequencies ranging from 1~Hz to 1~kHz. 

As previously described~\cite{pottierPRL2015}, the laser is used for three different purposes. First, it induces a thermocapillary flow in the liquid film, so that the thickness decreases locally. Second, the laser reflections at the air/liquid and liquid/solid interface interfere and are collected on the photodiode, allowing accurate measurements of the film thickness by interferometry. Lastly, the laser is used to measure the fluctuations of the slope of the liquid layer interface, which result in fluctuations of the position of the reflected beam.

We first focus on the thinning of the oil film due to the thermocapillary effect. The colored glass substrate is slightly heated by the laser beam. The temperature of the film therefore increases by thermal diffusion within a few seconds. Convection is negligible in  the liquid phase since the thermal P\'eclet number is much smaller than unity. The resulting surface tension gradient creates a liquid  flow directed away from the laser beam, which induces a local thinning of the film (see Fig.~\ref{fig1}). Since the absorption length of the glass substrate is very large compared to its thickness ($=2.5$~mm), it is the latter millimetric length that sets the lateral scale of the thermal gradient, and hence of film thinning. The extension of the thin flat region is therefore three orders of magnitude larger than the actual beam size ($R=1.5~\mu$m). 

To describe the thinning of the film, we denote  $\theta(r)=T(r)-T_0$ the temperature rise, with $T_0$ the room temperature and $r$  the radial coordinate  with origin at the beam center. The temperature rise being small ($<1$~K, see below), we neglect the variations of the physical properties of the liquid  that are at most 1\% of their equilibrium values. The temperature dependance of the surface tension is only accounted for  in the thermocapillary stress ($\approx 0.1$~Pa) resulting from the thermal gradient ($\approx 10^3$~K$\cdot$m$^{-1}$).
The film thickness $h(r,t)$ then follows the thin-film equation~\cite{oronRMP1997,supplmat}
%%%%%%%%%%%%%%%%%%%%%%%%%%%%%%%%%%%%%%%%%%%%%%%%%%%%%%%%
\begin{equation}
\partial_t  h = \frac{1}{3\eta r} \partial_r  rh^3 \left\{ \partial_r \left[ - \gamma_0 \nabla^2  h +  \phi'(h)   \right] +\frac{3\gamma_{\theta}}{2h}  \frac{\d \theta}{\d r} \right\}  \ ,
\label{thinfilmdyn}
\end{equation}
%%%%%%%%%%%%%%%%%%%%%%%%%%%%%%%%%%%%%%%%%%%%%%%%%%%%%%%%
with $\eta$ the liquid viscosity, $\gamma_0$ the equilibrium surface tension, and $\gamma_{\theta}=\vert \partial \gamma/ \partial \theta \vert$ the derivative of surface tension with respect to temperature. Since we focus on sub-micron scales, gravity can safely be neglected. The first term in parentheses  corresponds to the Laplace pressure. The second term $\phi'(h)$ is the derivative of the van der Waals interaction potential  (per unit area); it is the opposite of the so-called disjoining pressure. Consistently with the literature on oil films~\cite{churaevCollJ2003}, we assume $\phi'(h)=A/(6\pi h^3 )$ with a negative Hamaker constant $A<0$. Both intermolecular and Laplace forces oppose the thinning of the film, which is driven by the thermocapillary stress --- the third term  in parentheses.

%%%%%%%%%%%%%%%%%%%%%%%%%%%%%%%%%%%%%%%
\begin{figure}
\includegraphics[width=0.98\columnwidth]{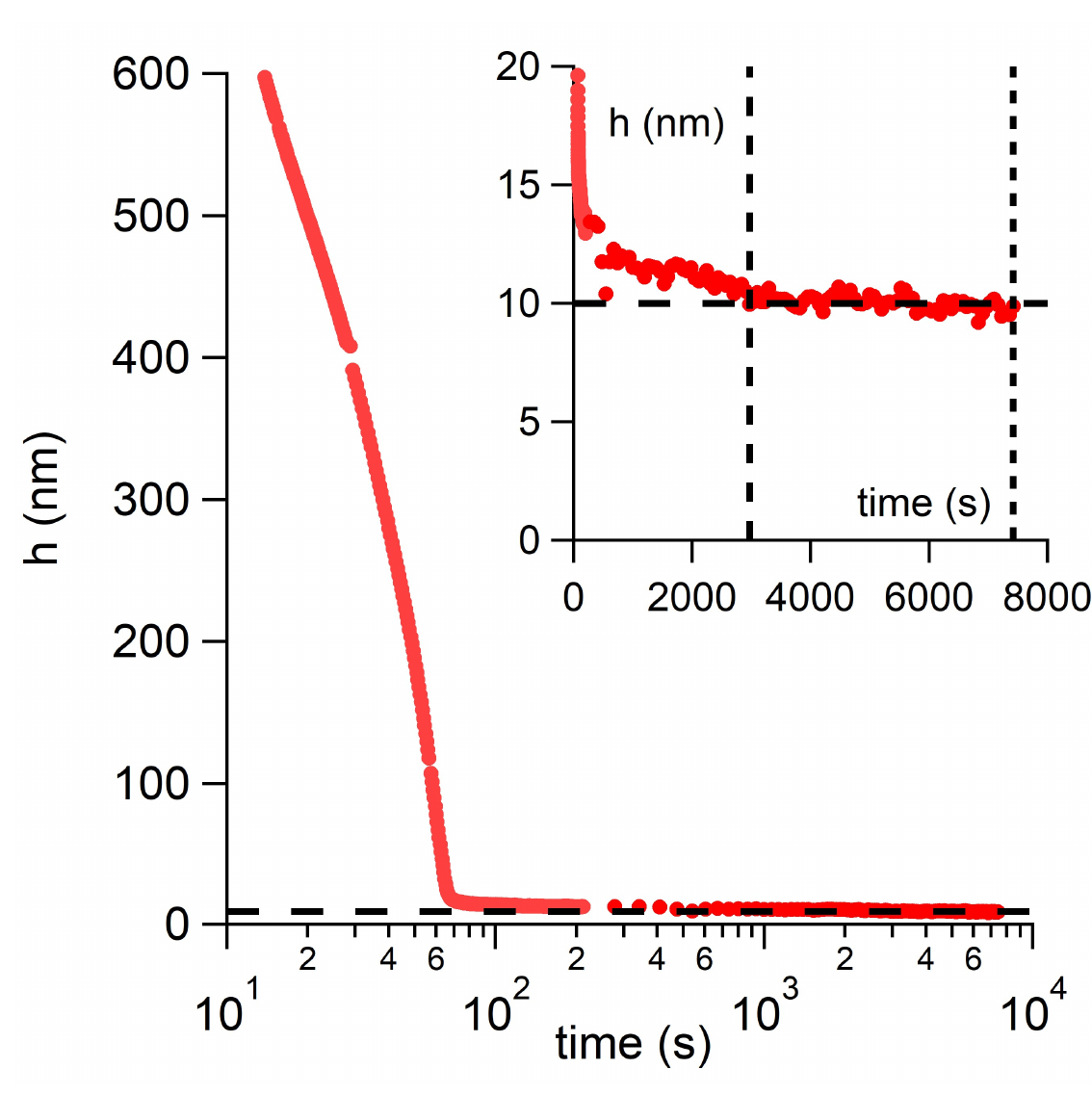}
\caption{Thickness of film of a silicone oil (viscosity $9.3$~mPa$\cdot$s) under the laser beam as a function of time. Insert shows details of the plateau corresponding to $h_0=10.0\pm 0.3$~nm. }
\label{fig2}
\end{figure}
%%%%%%%%%%%%%%%%%%%%%%%%%%%%%%%%%%%%%%%

A comprehensive study of the film thinning kinetics will be detailed in a future work. We focus here on the stationary version of Eq.~(\ref{thinfilmdyn}). Neglecting the Laplace pressure term, a double integration of Eq.~(\ref{thinfilmdyn}) yields  the stationary thickness under the beam $h_0=h(r=0)$~\cite{supplmat}
%%%%%%%%%%%%%%%%%%%%%%%%%%%%%%%%%%%%%%%%%%%%%%%%%%%%%%%%
\begin{equation}
h_0 = \left(  \frac{\vert A\vert }{6 \pi \gamma_{\theta} \theta(0)} \right)^{1/2} \ ,
\label{thickness}
\end{equation}
%%%%%%%%%%%%%%%%%%%%%%%%%%%%%%%%%%%%%%%%%%%%%%%%%%%%%%%%
where $\theta(0)$ is the maximum excess temperature. The  thickness  is therefore set by~$\theta (0)$ but is independent of both the initial film thickness, the viscosity and the details of the temperature profile.
With $\gamma_{\theta}=6.5\times10^{-5}$~N$\cdot$m$^{-1}\cdot$K$^{-1}$~\cite{ricciSCT1986} and taking for instance $\theta(0)=1$~K,  we predict  a stationary thickness $h_0 \approx 9$~nm for  $\vert A \vert=10^{-19}$~J.

We show in Fig.~\ref{fig2} an exemple of  the measured variation of an oil film thickness.  After an initial decrease, the stationary state  is reached within a few minutes. The thickness value at the plateau $h_0=10.0\pm 0.3$~nm remains remarkably stable for several hours.
Let us emphasize that the liquid plateau is very flat since it actually extends over the millimetric size of the thermal gradient.  The original Eq.~(\ref{thickness}) expresses the balance between intermolecular and thermocapillary forces. From this perspective, our experiment is a novel analogous to the Thin Film Pressure Balance, in which the disjoining pressure of a liquid film is measured at equilibrium with an imposed pressure~\cite{saramagoCOCIS2010}. Here, the film thickness is simply controlled by adjusting the intensity of the laser beam. Using a variable neutral density, we can check that thickness varies as predicted with laser intensity (see Fig.~S1 in the Supplemental Material~\cite{supplmat}). Stationary thicknesses ranging from 8 to 30~nm are obtained by tuning the laser intensity.

 In the following, we take advantage of the long-lasting stability of the thin films to measure fluctuation spectra over a wide frequency range. We especially focus  on low frequencies that are most sensitive to intermolecular forces.
As already mentioned, the same laser is used to measure interfacial fluctuations. More precisely, it is the fluctuating slope of the liquid surface that results in fluctuations in the position of the specularly reflected beam off the surface. In contrast to scattering experiments, in which a single surface mode is selected, the contributions of all the surface modes are measured but with different weights. Actually, the finite radius $R$ of the beam sets the lenghtscale of the measurement and the main contribution to the signal comes from the fluctuating modes with wavevectors $q \sim 1/ R$~\cite{tayRSI2008}. The power spectral density (PSD) can then be quantitatively expressed  as $S(\omega) \propto \int_0^{\infty}P(q,\omega) \Phi(q) \d q$ --- see Ref.~\cite{pottierPRL2015} for  a comprehensive discussion of $S(\omega)$ and all the parameters on which it depends. The weight function $\Phi(q)$ only depends on the beam characteristics (size and divergence), and is maximum for $q = 2\pi / R$. The mode density  $P(q,\omega)$ represents the PSD of the liquid film: $P(q,\omega)=\langle   \vert \tilde{h}_{\mathbf{q},\omega} \vert^2\rangle$, with $ \tilde{h}_{\mathbf{q},\omega}$ the Fourier transform of the height $h(\mathbf{r},t)$.

The mode density can be obtained in the lubrication  regime  $q h_0 \ll 1$, with $h_0$ a few tens of nanometers. Note that the lubrication theory cannot always be used  to describe thin film dynamics: for instance, it has been shown recently to fail to describe the very last moments before film rupture~\cite{morenoPRFluids2020}. 
Here however, the weight function $\Phi(q)$ is maximum for $q\sim 1/R$, with $R=1.5~\mu \text{m} \gg h_0$. The lubrication approximation is therefore fully legitimate (see SM for the derivation of the fluctuation spectra beyond the lubrication approximation~\cite{supplmat}). 
Our derivation of the  stochastic version of the thin film Eq.~(\ref{thinfilmdyn}) then follows Ref.~\cite{meckeJPCM2005}. Assuming small deviations with respect to the stationary height $h_0$, we can write $h(\mathbf{r},t)=h_0+ \delta h(\mathbf{r},t)$ with $\vert \delta h \vert \ll h_0$. The thin film equation can then be linearized and one obtains at lowest order~\cite{supplmat}
%%%%%%%%%%%%%%%%%%%%%%%%%%%%%%%%%%%%%%%%%%%%%%%%%%%%%%%%
\begin{equation}
\partial_t \delta h = \frac{h_0^3}{3\eta} \left[   \phi''(h_0) \nabla^2 \delta h - \gamma_0 \nabla^4 \delta h \right] + \kappa \bm{\nabla} \cdot \mathbf{f} \ ,
\label{thinfilmstoch}
\end{equation}
%%%%%%%%%%%%%%%%%%%%%%%%%%%%%%%%%%%%%%%%%%%%%%%%%%%%%%%%
where the stochastic term $\mathbf{f}=(f_x,f_y)$ follows a Gaussian distribution with ensemble average $\langle f_i (\mathbf{r},t) \rangle =0$ and $\langle f_i (\mathbf{r},t)f_j (\mathbf{r}',t')\rangle = \delta_{ij} \delta(\mathbf{r}-\mathbf{r}')\delta (t-t')$. The noise amplitude is $\kappa=\sqrt{2k_BT_0 h_0^3/(3\eta})$. It is important to note that the  thermocapillary stress happens to be irrelevant as far as  fluctuations are concerned. Indeed, although it sets the stationary thickness of the film, its contribution to the fluctuation spectrum is several orders of magnitude smaller than the capillary stress and the disjoining pressure~\cite{supplmat}. It is then straightforward to show that the mode density $P(q,\omega)$ is a Lorentzian function of the frequency
%%%%%%%%%%%%%%%%%%%%%%%%%%%%%%%%%%%%%%%%%%%%%%%%%%%%%%%%
\begin{equation}
P(q ,\omega) =  \frac{k_BT_0}{\pi}  \frac{\Gamma (q,h_0)}{\omega^2 + \omega_q^2} \  .
\label{correlthin}
\end{equation}
%%%%%%%%%%%%%%%%%%%%%%%%%%%%%%%%%%%%%%%%%%%%%%%%%%%%%%%%
The relaxation rate is given by  $\omega_q=\frac{\gamma_0}{\lambda^{2}} \Gamma(q,h_0 )(1+q^2\lambda^2)$, with $\Gamma(q,h_0 )=q^2 h_0^3/(3\eta)$ the dissipation kernel.  
Since gravity is negligible, the capillary length is replaced by $\lambda= \sqrt{ 2\pi h_0^4 \gamma_0/\vert A \vert}$. This lengthscale lies in the micrometer range for typical values $\vert A \vert \approx10^{-19}$~J, $\gamma_0 \approx 10^{-2}$~N$\cdot$m$^{-1}$, and $h_0 \approx 10^{-8}$~m. As a consequence, the contributions from both capillary and intermolecular forces are equally significant in the PSD. Interestingly, $\lambda$ also corresponds to the critical wavelength of the unstable modes leading to the rupture of metastable films~\cite{vrijFaraday1966,meckeJPCM2005}. Eq.~(\ref{correlthin}) moreover reveals that the main contribution of van der Waals forces is to shift the low cut-off frequency of the Lorentzian function. Indeed, no modification of the capillary spectrum is expected at high frequencies, where $P(q,\omega)\sim \omega^{-2}$. In contrast, one expects a signature of van der Waals forces on the fluctuation spectrum at low frequencies, below  a few tens of Hz taking the same numerical values as above.

%%%%%%%%%%%%%%%%%%%%%%%%%%%%%%%%%%%%%%%
\begin{figure}
\includegraphics[width=\columnwidth]{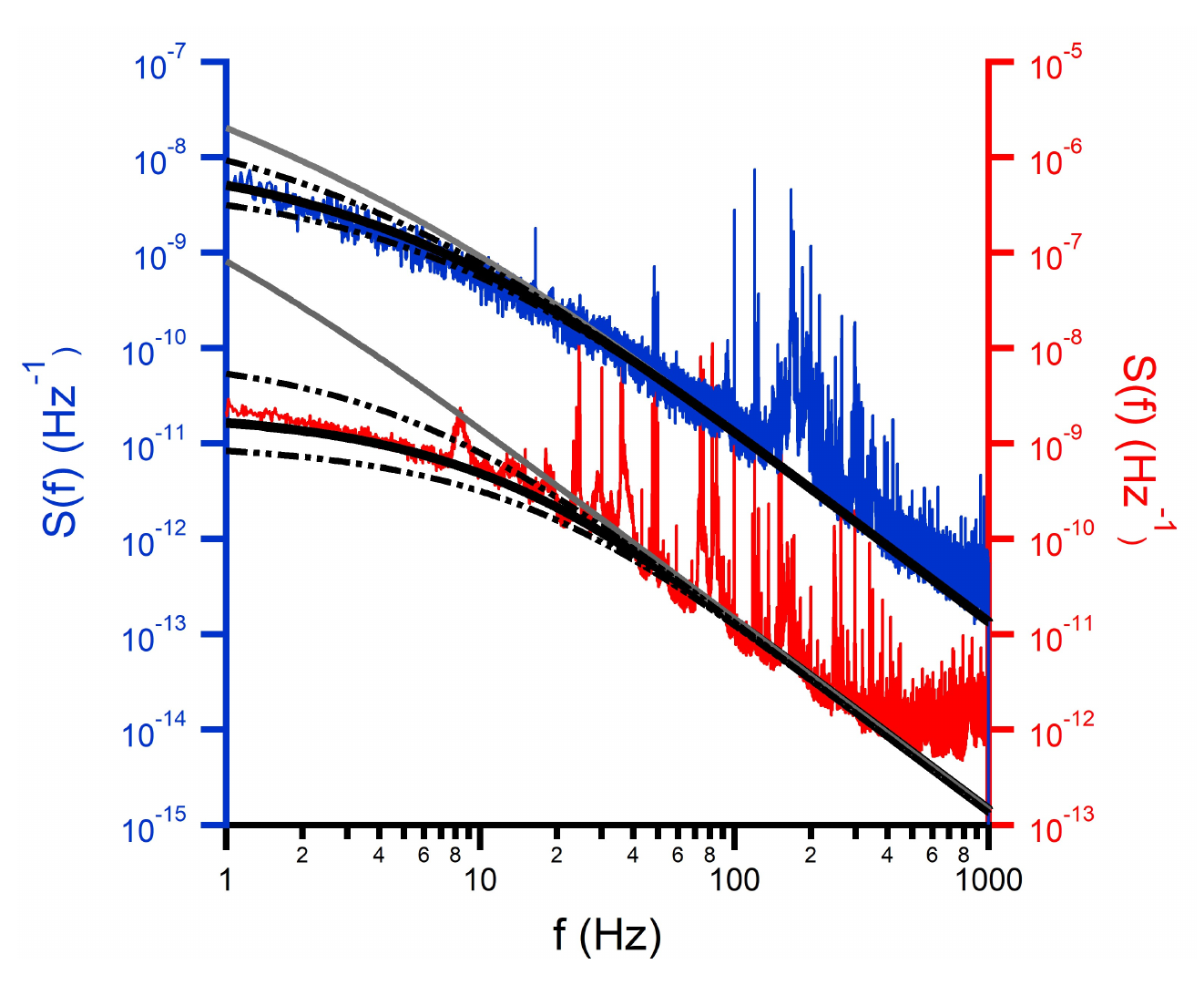}
\caption{Fluctuation spectra of oil films of thickness $30.3$~nm (viscosity $19$~mPa$\cdot$s, blue, left axis) and $12.9$~nm (viscosity $9.3$~mPa$\cdot$s, red, right axis). Spurious peaks result from parasitic noise either of mechanical or electromagnetical origin. For each spectrum, the full lines represent the theoretical spectra without (grey lines) and with (black lines) intermolecular forces effects. Dash-dotted lines correspond to spectra computed with Hamaker constants differing by $\pm 50 \%$.}
\label{fig3}
\end{figure}
%%%%%%%%%%%%%%%%%%%%%%%%%%%%%%%%%%%%%%%

We  thus anticipate a significant contribution of intermolecular forces  for thicknesses smaller than a few tens of nanometers and at low frequencies. Fig.~\ref{fig3} displays the experimental spectra of two different oil films with two different thicknesses: $12.9$~nm (viscosity $9.3$~mPa$\cdot$s)  and $30.3$~nm (viscosity $19$~mPa$\cdot$s). The data are obtained in the stationary regime and averaged over at least 20 spectra (the acquisition time of each spectrum being 50~s). The experimental spectra are in excellent agreement with the theoretical prediction obtained by numerical integration of the mode density Eq.~(\ref{correlthin}). 
Tabulated values of viscosity, surface tension, and refractive indexes were used, together with measured values of film height and beam radius. The only adjusted parameter is therefore the Hamaker constant: we get $\vert A \vert = 1.2\times10^{-19}$~J and $\vert A \vert = 2\times 10^{-19}$~J for the less and more viscous oil, respectively. 
For comparison, the theoretical spectra without the contribution of intermolecular forces are also shown. Clearly, it is the low-frequency part of spectra which is most strongly modified by van der Waals forces. 
One can notice that the Hamaker constants for silicon oils on glass are one order of magnitude larger than for alkane films on mica, but they are very close to alkane films on steel~\cite{churaevCollJ2003}. Silicone oils being known to be particularly difficult to remove from solid surfaces, this finding consistently indicates a strong repulsion between glass/oil and oil/air surfaces.

%%%%%%%%%%%%%%%%%%%%%%%%%%%%%%%%%%%%%%%
\begin{figure}[b]
\includegraphics[width=0.7\columnwidth]{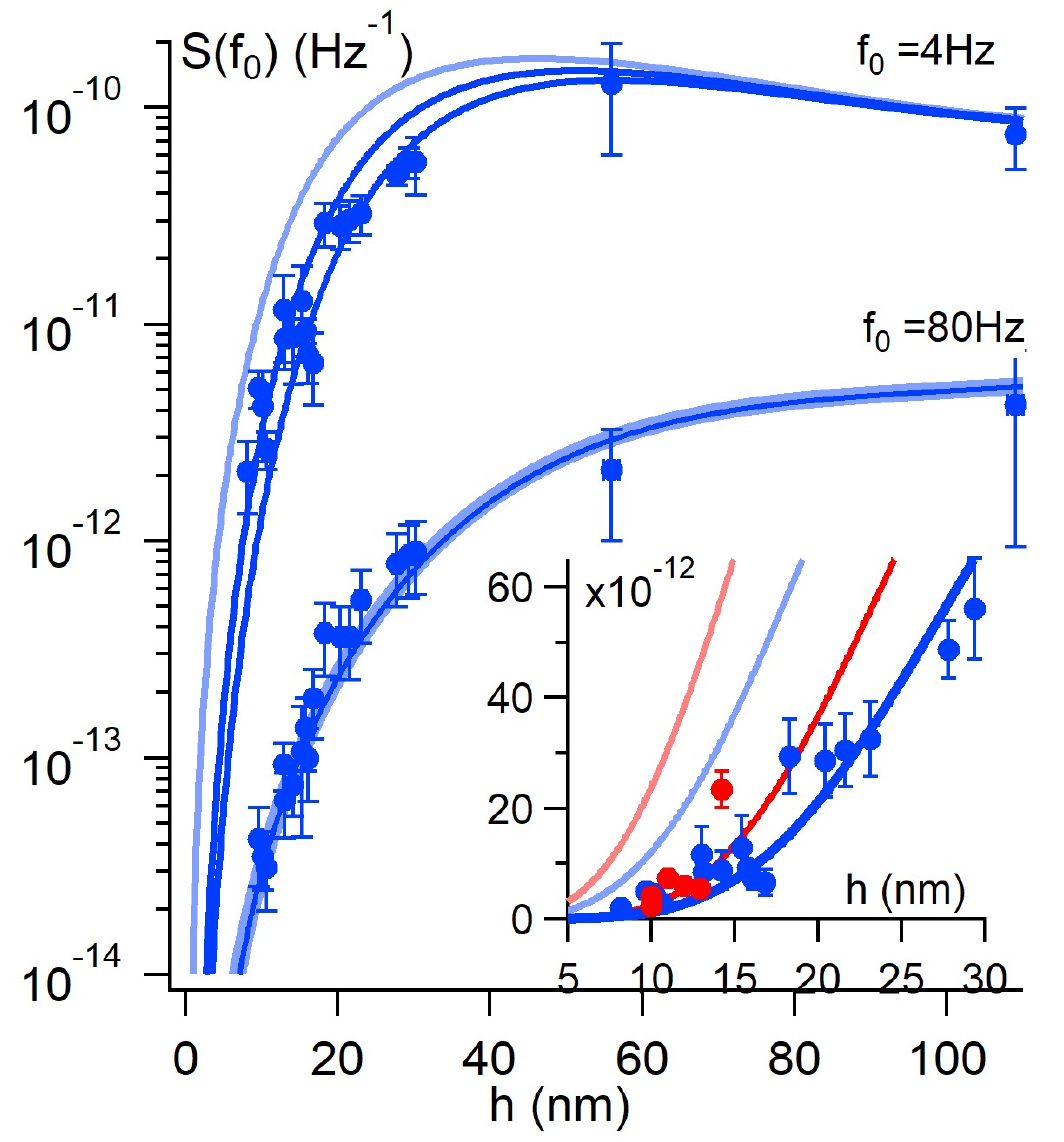}
\caption{Fluctuation spectra (in logarithmic scale) of  oil films (viscosity $19$~mPa$\cdot$s) at $4$~Hz and $8$0~Hz, as a function of film thickness. The full lines correspond to theoretical predictions without (lighter lines) and with (darker lines) intermolecular forces, with $\vert A \vert =10^{-19}$~J and $\vert A \vert = 2\times10^{-19}$~J.  Inset: Fluctuation spectra (in linear scale) at $4$~Hz  of the two oil films with viscosity $19$~mPa$\cdot$s (blue circles, same data as main figure) and $9.3$~mPa$\cdot$s (red circles). Vertical error bars correspond to the statistical uncertainty. Horizontal error bars are too small to appear.}
\label{fig4}
\end{figure}
%%%%%%%%%%%%%%%%%%%%%%%%%%%%%%%%%%%%%%%

Additional experiments are performed where the film thickness is varied by adjusting the laser intensity. In each experiment, the fluctuation spectra of a stationary thin film are measured at two different frequencies, respectively 4~Hz and 80~Hz. The corresponding values of the PSD, respectively averaged over a $0.8$~Hz-wide and $16$~Hz-wide frequency range, are reported in Fig.~\ref{fig4} as a function of the film thickness. Note that for the thicker films, measurements are conducted before reaching the plateau as the film is thinning down. Only three spectra could then be recorded and the large error bars result from poor averaging. All data points are in excellent agreement with our theoretical predictions. Fig.~\ref{fig4} confirms that the effect of intermolecular forces is most significant for the lower frequencies and for smaller thicknesses. In contrast, intermolecular forces do not modify the spectra at the larger frequencies, even for film thicknesses below 10 nm. Similar data are obtained for the two different oils. Note that the Hamaker constants are expected to be very close
since optical indexes of the two silicon oils only differ by 0.1\%.

To summarize,  we have developed a novel method to form oil films of tunable nanometric thicknesses using a laser-induced thermocapillary effect. The films are stable for several hours, allowing for the characterization of the fluctuation dynamics of their interface.  We thus present the first evidence of the contribution of intermolecular forces on the fluctuation spectra of supported liquid films, over a large frequency range. The experimental data are fully described by our theoretical predictions accounting for van der Waals forces, the Hamaker constant being the only adjusted parameter. A significant decrease of the amplitude of the spectra is observed at low frequencies, showing the relevance of intermolecular forces in nanometric liquid layers.  Our results therefore provides new insight on the dynamics of ultra-thin liquid films.

\acknowledgments

This project has received funding from the European Union’s Horizon 2020 research and innovation program under the Marie Sklodowska-Curie grant agreement No 754387.

\end{document}